\documentstyle[twoside,fleqn,my_espcrc2,epsf]{article}

\newcommand{\ewxy}[2]{\setlength{\epsfxsize}{#2}\epsfbox[10 60 640 570]{#1}}

\newcommand{\be}{\begin{equation}}
\newcommand{\ee}{\end{equation}}

\newcommand{\fig}[1]{Fig.~\ref{#1}}
\newcommand{\eq}[1]{Eq.~(\ref{#1})}

\newcommand{\Gev}{{GeV}}
\newcommand{\Mev}{{MeV}}
\newcommand{\bbbar}{{b\bar b}}
\newcommand{\ccbar}{{c\bar c}}
\newcommand{\dm}{{\delta M}}
\newcommand{\de}{{\delta E}}

\newcommand{\mv}{{M_V}}
\newcommand{\mps}{{M_{PS}}}

\newcommand{\ie}{{\rm i.e.\ }}

\newcommand{\ainv}{a^{-1}}

\newcommand{\AmS}{{\protect\the\textfont2
  A\kern-.1667em\lower.5ex\hbox{M}\kern-.125emS}}

\title{$\Upsilon$ and $J/\Psi$ spectroscopy using clover fermions
in the presence of dynamical quarks}

\author{S. Collins, R.G. Edwards, U.M. Heller, and 
J. Sloan\address{SCRI, Florida State University}
}

\begin{document}

\begin{abstract}
We calculate spin average splittings in the $\bbbar$ and $\ccbar$
systems using the clover action.  We compare static and kinetic
masses of heavy-heavy and heavy-light mesons and discuss the
consistency of the results.
\end{abstract}

\maketitle

\section{INTRODUCTION}

In this poster presentation, we discuss recent preliminary
calculations of spectroscopy in the $\bbbar$ and $\ccbar$ systems,
which were carried out using the tadpole improved Clover action and
interpreted in the limit of heavy quarks using the Fermilab
interpretation of this action \cite{fnal}.  In the first half, we
discuss our results for the spin-averaged $1P-1S$ and $2S-1S$
splittings, which are commonly used to extract
$\alpha_{\overline{MS}}(M_Z)$, and compare to those of other groups.  In
the second half, we investigate the relationship between the static
and kinetic masses of mesons, $M_1$ and $M_2$ respectively, composed
of various mixtures of `heavy' and `light' quarks (\ie heavy-heavy,
heavy-light, and light-light). The Fermilab interpretation of the Clover 
action with temporal and spatial $\kappa$ values equal 
(i.e.\ {\it  symmetric $\kappa$}) identifies
the kinetic mass as the physical mass and differences between kinetic and 
static masses
as a shift in the energy zero of the quarks.  This interpretation requires
\be
\dm(HH) + \dm(LL) = 2\dm(HL),
\ee
where $\dm = M_2 - M_1$, and HH, HL, and LL are different flavor
combinations of heavy and light quarks.  We find that this is obeyed,
at this particular value of $n_f$ and $\beta$, for $M_2(HL) < 1$, but
is strongly violated for $M_2(HL) >1$.  The implication of this is
that, for example, there is no choice of $\kappa_b$ (for our ensemble)
for which $M_2(\Upsilon)$ and $M_2(B)$ simultaneously agree with
experiment\footnote{This is due to the large dimensionless mass, {\it not}
quenching effects on the running of the strong coupling}.

\section{THE SIMULATION AND FITTING}
The gauge configurations, smearing procedures, and fitting procedures are
discussed in \cite{clov_wils}.  For $P$ states, an $l=1$ bound state spatial 
smearing function was used; one of the quarks was 
smeared by a gaussian multiplied by $x+iy$, where $x$ and $y$ are
the values of the $x$ and $y$ coordinates, and the other quark was a point 
source.  For the ``local'' $P$-wave sink we used similar smearings 
which combined derivatives of a delta function.
We used 
three different smearing radii each for $\bbbar$ and $\ccbar$, and 
performed both matrix and vector fits with one, two, and three exponentials.
All fits were done with $t_{max}=15$, i.e.\ ommiting the central point of
the propagator.

Because onia (especially $\Upsilon$) are so small, our physical box contained
several relatively decorrelated volumes.  We took advantage of this by 
superposing the quark propagators from 8 spatial origins, using $Z(3)$ noise as 
described in \cite{znoise}.  For $P$ states, we further
doubled our statistics by using $t=16$ (in addition to $t=0$) as an initial
timeslice.  This was vital in obtaining a reliable $1P$-$1S$ splitting; the
effective mass plots with the original statistics showed a ``bump'' near
the center which led to ambiguous fits.

For our determination of kinetic masses, we fit the ratio of a
momentum one state and a zero momentum state to a single exponential, using 
a plateau of 8-12
in all cases.  Looking at effective mass plots of the ratio, we saw
that this corresponded to a very conservative value for $t_{min}$, so
our statistical errors should be much larger than the (systematic)
fitting errors.

\section{SPECTROSCOPY RESULTS}

\begin{table*}[thb]
\caption{Selected $\bbbar$ and $\ccbar$ fit results.}
\label{tab:ups_results}
\begin{tabular}{c|cccccc|ccc}
\hline
State & type & Smears & $N_{cosh}$ & $T_{min}$& $T_{max}$ &Q & $E_1$ & $E_2$ & $E_3$ \\
\hline
$\eta_b(nS)$
      & Vec  &  a,b,c   & 3 &   5 &15& .06 & 2.4539(8) & 2.709(15) & 2.94(3) \\
      & Vec  &  a,b     & 2 &  10 &15& .37 & 2.4532(9) & 2.763(17) &         \\
      & Vec  &  b,c     & 3 &   2 &15& .03 & 2.4539(7) & 2.690( 7) & 3.19(6) \\
      & Mat  &  a,b     & 3 &   3 &15& .11 & 2.4543(7) & 2.70 (2 ) & 2.96(4) \\
      & Mat  &  a,b     & 2 &   8 &15& .18 & 2.4545(6) & 2.751( 7) &         \\
      & Mat  &  a,b     & 2 &  11 &15& .72 & 2.4533(6) & 2.701(15) &         \\
\hline
$\Upsilon(nS)$
      & Vec  &  a,b,c   & 3 &   5 &15& .07 & 2.4707( 9) & 2.714(15) & 2.95(3) \\
      & Vec  &  a,b     & 3 &   2 &15& .29 & 2.4691(13) & 2.76 (3 ) &         \\
      & Mat  &  a,b     & 3 &   3 &15& .13 & 2.4710( 8) & 2.71 (2 ) & 2.97(4) \\
\hline
$h_b(nP)$
      & Vec  &  a,b,c   & 3 &   1 &15& .30 & 2.666(7) & 2.87(2) & 3.083(19)\\
      & Vec  &  a,b,c   & 2 &   7 &15& .41 & 2.668(9) & 2.95(3) &          \\
      & Mat  &  a,b,c   & 2 &   9 &15& .34 & 2.669(8) & 2.94(4) &          \\
\hline
%
$\eta_c(nS)$
      & Vec  &  a,b,c   & 3 &   6 &15& .24 & 1.2156(17) & 1.48(8) & 1.78(15) \\
      & Vec  &  a,b,c   & 2 &   9 &15& .56 & 1.2150(13) & 1.53(3) &          \\
      & Vec  &  a,b     & 2 &   8 &15& .55 & 1.2157(12) & 1.55(2) &          \\
      & Mat  &  a,b     & 3 &   3 &15& .14 & 1.2137(15) & 1.47(7) & 1.77(11) \\
      & Mat  &  a,b     & 2 &   9 &15& .45 & 1.2138(13) & 1.49(3) &          \\
\hline
$J/\psi(nS)$
      & Vec  &  a,b,c   & 3 &   4 &15& .20 & 1.2547(15) & 1.53(2) & 1.96(4)  \\
      & Vec  &  a,b     & 3 &   4 &15& .16 & 1.2541(2 ) & 1.51(4) & 1.86(3)  \\
\hline
$h_c(nP)$
      & Vec  &  a,b,c   & 3 &   2 &15& .66 & 1.481( 8)  & 1.86(3)   & 2.52(2) \\
      & Vec  &  a,b,c   & 2 &   7 &15& .63 & 1.459(18)  & 1.79(7)   &         \\
      & Vec  &  a,b     & 2 &   2 &15& .52 & 1.490( 5)  & 1.863(10) &         \\
      & Mat  &  a,b     & 2 &   5 &15& .27 & 1.484( 9)  & 1.79(4)   &         \\
      & Mat  &  a,c     & 2 &   6 &15& .14 & 1.469( 8)  & 1.84(2)   &         \\
\hline
\end{tabular}
\end{table*}

We present the results of a representative sample of our fits in 
Table~\ref{tab:ups_results}.
The $\bbbar$ fits seemed
better behaved than the $\ccbar$ fits; this can be seen by comparing
$h_b$ to $h_c$ results. 
We see that corresponding $\Upsilon$ and $\eta_b$ fits are very similar;
this is also true of the $J/\psi$ and $\eta_c$.  
We quote preliminary splittings in Table~\ref{tab:splits}; the error
bars include our estimate of sytematic fitting errors.  This leads to
inverse lattice spacings of $2.27(13)$ and $2.17(25) \Gev$ from the
bottomonium $1P$-$1S$ and $2S$-$1S$ splittings, respectively, and
$2.0(2)$ and $2.3(5) \Gev$ from charmonium.  For the
respective dimensionless kinetic masses, $aM_2$, of the lowest energy vector
states we find $4.45(6)$ and $1.44(3)$ (corresponding to dimensionful
masses of $10.1(6) \Gev$ for ``$\Upsilon$'' and $2.9(3) \Gev$ for ``$J/\Psi$''
when combined with the corresponding $1P$-$1S$ $\ainv$.)
Hyperfine splittings were estimated by examining effective mass plots
of the ratio of vector to pseudo-scalar propagators; rough values
of $41(5) \Mev$ for bottomonium and $80(20) \Mev$ for charmonium were 
obtained.

\begin{table}[thb]
\vskip -7mm
\caption{Preliminary splittings}
\label{tab:splits}
\begin{tabular}{c|c}
\hline
$h_c(1P)-\ccbar(\overline{1S})$ & 0.23(2)  \\\hline
$J/\psi(2S)-J/\psi(1S)$ & 0.26(5) \\\hline
$h_b(1P)-\bbbar(\overline{1S})$ & 0.199(11) \\\hline
$\Upsilon(2S)-\Upsilon(1S)$ & 0.26(3) \\\hline
\end{tabular}
\vskip -5mm
\end{table}

NRQCD $\bbbar$ and Wilson $\ccbar$ spectroscopy has been studied 
on this ensemble
by other groups~\cite{nrdyn,wdch}.  The NRQCD inverse lattice
spacings from the $\bbbar$ $1P$-$1S$ and $2S$-$1S$ splittings are
$2.44(7)$ and $2.37(10)$, respectively, roughly $10\%$ higher than our
values.  This is consistent with experience from quenched calculations,
where NRQCD yields splittings about $10\%$ smaller than corresponding
clover calculations~\cite{sloan94}.  Our $\ccbar$ $1P$-$1S$ inverse 
lattice spacing is consistent with, but larger than that quoted for
Wilson fermions in~\cite{wdch}. Because the clover term mainly affects 
spin-dependent splittings, we expect agreement between Wilson and clover 
calculations of the spin-averaged spectrum.

\section{KINETIC MASSES}

QCD is a Lorentz invariant theory. Therefore the dispersion relation
of physical states should obey the relation $E(p^2) = \sqrt{p^2+M^2}$.
In~\cite{fnal} it was pointed out that, for non-relativistic particles,
a discretization which subtracts off a zero of energy for each (heavy)
quark captures the essential physics:
\be
E(p^2) - \sum_i{\dm_i} = \sqrt{p^2+M^2} - \sum_i{\dm_i},
\label{eq:shifted_disp}
\ee
where the lhs is typically called the static mass, $M_1$, and $\dm_i$ is 
the shift in the zero of energy of the $i$'th quark in the state.
The kinetic mass is defined as $M_2 = (2\frac{dE(p^2)}{dp^2}|_{p^2=0})^{-1}$,
and is equal to $M$ for the dispersion relation~(\ref{eq:shifted_disp}).
A useful definition of $M_2$
given the splitting $\de = E(p^2)-E(0)$ is obtained by solving 
\eq{eq:shifted_disp} for $M$; one obtains $M_2 = (p^2/2\de) - \de/2$.

The sum over energy shifts in \eq{eq:shifted_disp} can be calculated in
simulations; it is just the discrepancy between kinetic and static masses
$\dm(state) = M_2(state) - M_1(state)$.  
Because the energy shifts in \eq{eq:shifted_disp} depend only on the
quark content of the state, states with the same quark content should
have the same $\dm$; the energy shift associated with a $B$ meson
should be the average of the $\Upsilon$ and $\pi$ shifts, i.e. half
the energy shift of the $\Upsilon$.  This leads us to the relation
$2\dm(HL) = \dm(HH) + \dm(LL)$ as a non-perturbative test of the 
consistency of the idea of attributing discrepancies between $M_1$ and
$M_2$ to a shift in the zero of energy. 

\begin{figure}[htb]
\vskip -17mm
\centerline{\ewxy{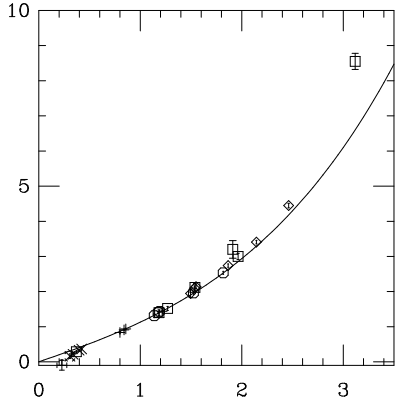}{80mm}}
\caption{
$M_2(PS)$ vs $M_1(PS)$ for HH and HL mesons. The perturbative curve for
HH mesons is shown as a solid line.
}
\label{fig:m2vsm1}
\vskip -5mm
\end{figure}

The energy shifts $\dm_i$ have been calculated for free quarks
in~\cite{fnal}; this leads to the relation
\be
M_{2,pert}(M_1) = \frac{e^{M_1}\sinh (M_1)}{\sinh (M_1) + 1}
\label{eq:m2pert}
\ee
In \fig{fig:m2vsm1}, we have plotted our results for $M_2(PS)$ 
vs.\ $M_1(PS)$
for both mesons composed of degenerate quarks (HH) and those of
mixed mass(HL).
It is striking that both HH and HL mesons appear to lie on a 
universal curve for this range of mass values.  We also include the 
`perturbative' curve for HH mesons $M_2 = 2M_{2,pert}(M_1/2)$,
which seems to describe the curve fairly well.  Indications of this
behavior were first seen in~\cite{collins93}, albeit with large error bars.

\begin{figure}[htb]
\vskip -17mm
\centerline{\ewxy{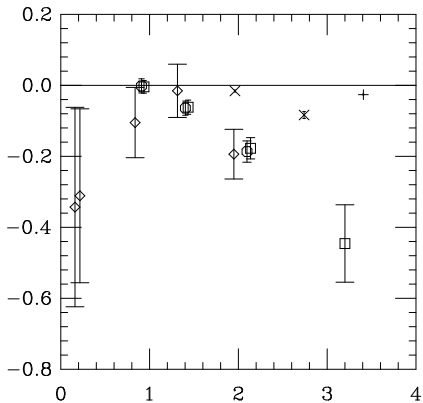}{80mm}}
\caption{$I(H,L)$ vs $M_2(HL)$.  Points with the same symbol have
the same lighter quark mass.
}
\label{fig:inconsistency}
\vskip -5mm
\end{figure}

In \fig{fig:inconsistency} we plot the relative inconsistency
\be
I(H,L) = \frac{ \dm(HL) - (\dm(HH)+\dm(LL))/2 }{M_2(HL)}
\label{eq:inconsistency}
\ee
as a function of $M_2(HL)$ for many different heavy-light mesons.
For $M_2(HL) < 1$, we see that \eq{eq:shifted_disp} is consistent;
for $M_2(HL) > 1$, however the {\it relative} discrepancy in
energy shifts appears to be growing linearly in $M_2(HL)$ for fixed light
quark mass.  Since
the $D$ meson should have a dimensionless mass of about $.7$ on this
ensemble, the energy shift ansatz is consistent for charmed mesons.
The $B$ meson, however, should have a dimensionless mass of about $2.5$;
if we tune $\kappa_b$ by matching $M_2(\Upsilon)$ to experiment (as
we did), then we expect $M_2(B)$ to be much too light.  Recall
that we found $M_2(\Upsilon) = 10.1(6) \Gev$ for our choice of
$\kappa_b$ (i.e.\ we used a bare quark mass slightly too heavy). 
For the same choice of $\kappa_b$ we find $M_2(B) = 4.4(3) \Gev$ -
much lighter than the physical $B$ meson!

\section{CONCLUSIONS}

Our results for spin-averaged splittings of the $\bbbar$ and $\ccbar$
systems showed no surprises.  All gave inverse lattice spacings between
$2.0$ and $2.3 \Gev$, with the smallest inverse lattice spacings
coming from the softest splittings.  This is consistent with a partial removal
of quenching effects by the two flavors of dynamical fermions.  The
$\Upsilon$ inverse lattice spacings were roughly $10\%$ smaller than 
in corresponding NRQCD calculations, consistent with experience in
quenched calculations.

A problem arises, however, when we examine the relationship between
static and kinetic masses.  We find that, after tuning $\kappa_b$ so that
the kinetic mass of the $\Upsilon$ agrees with experiment, the kinetic mass
of the $B$ meson is roughly $20\%$ too small.  This will obviously
cause difficulties with trying to calculate $f_B$ directly at the $B$
meson mass on a $2 \Gev$ lattice, especially since $f_B$ is strongly
dependent upon $M_B$.  This problem does not appear to be present in
NRQCD; \eq{eq:inconsistency} has been checked for the $B_c$ 
system~\cite{nrqcdhl}.  Another area where this inconsistency may cause
problems is in low-$\beta$ calculations; $aM_\rho \approx 1.5$ when
the inverse lattice spacing is $500 \Mev$.  With this in mind, the dependence
of the inconsistency threshhold upon lattice spacing should be checked.

Recently, another group has postulated a particular form of the lattice
dispersion relation for mesons~\cite{lanl}, which leads them to similar
conclusions to those presented here.  We stress that we calculate both
$M_2$ and $M_1$ directly from the simulation, rather than relating them
through a particular ansatz for the dispersion relation.  

Finally, we would like to stress that our results do not prove that
$B$ physics cannot be done on a $2 \Gev$ lattice.  First, we know that
the clover action does not include retardation effects correctly for 
very heavy quarks; it is possible that adding improvement terms to correct
this would reduce the inconsistency.  A more promising idea is to
explicitly break the hypercubic symmetry of the clover action by using
different $\kappa$'s in the temporal and spatial directions~\cite{fnal}; 
the asymmetry could then be tuned to require that $M_1$ agree with $M_2$,
obviating the need for energy shifts.

\section*{ACKNOWLEDGEMENTS}
We thank Andreas Kronfeld for useful conversations.
The computations were performed on the CM-2 at SCRI.
This research was supported by DOE contracts DE-FG05-85ER250000 
and DE-FG05-92ER40742.

\end{document}